\documentclass[10pt,a4paper,onecolumn]{article}
\usepackage{marginnote}
\usepackage{graphicx}
\usepackage{xcolor}
\usepackage{authblk,etoolbox}
\usepackage{titlesec}
\usepackage{calc}
\usepackage{tikz}
\usepackage{hyperref}
\hypersetup{
    colorlinks=true,     
    linkcolor=blue,      
    citecolor=blue,      
    filecolor=blue,      
    urlcolor=blue        
}
\usepackage{caption}
\usepackage{tcolorbox}
\usepackage{amssymb,amsmath}
\usepackage{ifxetex,ifluatex}
\usepackage{seqsplit}

\usepackage[top=3.5cm, bottom=3cm, right=1.5cm, left=1.5cm,
            headheight=2.2cm, reversemp, includemp, marginparwidth=4.0cm]{geometry}

\titleformat{\section}
  {\normalfont\sffamily\Large\bfseries}
  {}{0pt}{}
\titleformat{\subsection}
  {\normalfont\sffamily\large\bfseries}
  {}{0pt}{}
\titleformat{\subsubsection}
  {\normalfont\sffamily\bfseries}
  {}{0pt}{}
\titleformat*{\paragraph}
  {\sffamily\normalsize}

\usepackage{fancyhdr}
\makeatletter
\let\ps@plain\ps@fancy
\fancyheadoffset[L]{4.5cm}
\fancyfootoffset[L]{4.5cm}

\definecolor{linky}{rgb}{0.0, 0.5, 1.0}

\newtcolorbox{repobox}
   {colback=red, colframe=red!75!black,
     boxrule=0.5pt, arc=2pt, left=6pt, right=6pt, top=3pt, bottom=3pt}

\newcommand{\ExternalLink}{%
   \tikz[x=1.2ex, y=1.2ex, baseline=-0.05ex]{%
       \begin{scope}[x=1ex, y=1ex]
           \clip (-0.1,-0.1)
               --++ (-0, 1.2)
               --++ (0.6, 0)
               --++ (0, -0.6)
               --++ (0.6, 0)
               --++ (0, -1);
           \path[draw,
               line width = 0.5,
               rounded corners=0.5]
               (0,0) rectangle (1,1);
       \end{scope}
       \path[draw, line width = 0.5] (0.5, 0.5)
           -- (1, 1);
       \path[draw, line width = 0.5] (0.6, 1)
           -- (1, 1) -- (1, 0.6);
       }
   }

\patchcmd{\@maketitle}{center}{flushleft}{}{}
\patchcmd{\@maketitle}{center}{flushleft}{}{}
\patchcmd{\@maketitle}{\LARGE}{\LARGE\sffamily}{}{}
\def\maketitle{{%
  
  \AB@maketitle}}
\makeatletter
\renewcommand\AB@affilnote[1]{{\bfseries #1}\hspace{3pt}}
\makeatother

\ifnum 0\ifxetex 1\fi\ifluatex 1\fi=0 
  \usepackage[T1]{fontenc}
  \usepackage[utf8]{inputenc}

\else 
  \ifxetex
    \usepackage{mathspec}
  \else
    \usepackage{fontspec}
  \fi
  \defaultfontfeatures{Ligatures=TeX,Scale=MatchLowercase}

\fi
\IfFileExists{upquote.sty}{\usepackage{upquote}}{}
\IfFileExists{microtype.sty}{%
\usepackage{microtype}
\UseMicrotypeSet[protrusion]{basicmath} 
}{}

\hypersetup{unicode=true,
            pdftitle={EspressoDB: A scientific database for managing high-performance computing workflows.},
            pdfborder={0 0 0},
            breaklinks=true}
\usepackage{graphicx,grffile}
\makeatletter
\def\maxwidth{\ifdim\Gin@nat@width>\linewidth\linewidth\else\Gin@nat@width\fi}
\def\maxheight{\ifdim\Gin@nat@height>\textheight\textheight\else\Gin@nat@height\fi}
\makeatother
\setkeys{Gin}{width=\maxwidth,height=\maxheight,keepaspectratio}
\IfFileExists{parskip.sty}{%
\usepackage{parskip}
}{
\setlength{\parindent}{0pt}
\setlength{\parskip}{6pt plus 2pt minus 1pt}
}
\providecommand{\tightlist}{%
  \setlength{\itemsep}{0pt}\setlength{\parskip}{0pt}}
\ifx\paragraph\undefined\else
\let\oldparagraph\paragraph
\renewcommand{\paragraph}[1]{\oldparagraph{#1}\mbox{}}
\fi
\ifx\subparagraph\undefined\else
\let\oldsubparagraph\subparagraph
\renewcommand{\subparagraph}[1]{\oldsubparagraph{#1}\mbox{}}
\fi

\title{EspressoDB: A scientific database for managing high-performance computing workflow}

        \author[1,2,3]{Chia Cheng Chang}
        \author[2,3$\dagger$]{Christopher K{\"o}rber}
        \author[3,2]{Andr{\'e}~Walker-Loud}

      \affil[1]{iTHEMS RIKEN, Wako, Saitama 351-0198, Japan}
      \affil[2]{Department of Physics, University of California, Berkeley, California 94720, USA}
      \affil[3]{Nuclear Science Division, Lawrence Berkeley National Laboratory, Berkeley, California 94720}
      \affil[$\dagger$]{\textit {\href{mailto:software@ckoerber.com}{software@ckoerber.com}}}

  \date{\vspace{-5ex}}

\begin{document}
\maketitle

\marginpar{
  \sffamily\small

  {\bfseries DOI:} \href{https://doi.org/10.21105/joss.02007}{\color{linky}{10.21105/joss.02007}}

  \vspace{2mm}

  {\bfseries Software}
  \begin{itemize}
    \setlength\itemsep{0em}
    \item \href{https://github.com/openjournals/joss-reviews/issues/2007}{\color{linky}{Review}} \ExternalLink
    \item \href{https://github.com/callat-qcd/espressodb}{\color{linky}{EspressoDB}} \ExternalLink
    \item \href{https://github.com/callat-qcd/lattedb}{\color{linky}{LattedDB}} \ExternalLink
    \item \href{https://doi.org/10.5281/zenodo.3677432}{\color{linky}{Archive}} \ExternalLink
  \end{itemize}

  \hrule

  \vspace{2mm}

  {\bfseries Editor:}
  
  \href{https://luc.edu/cs/people/ftfaculty/gkt.shtml}{\color{linky}{George K. Thiruvathukal}} \ExternalLink

  \vspace{2mm}

  {\bfseries Reviewers:}
  \begin{itemize}
    \setlength\itemsep{0em}
    \item \href{https://github.com/remram44}{\color{linky}{@remram44}}
    \item \href{https://github.com/ixjlyons}{\color{linky}{@ixjlyons}}
  \end{itemize}

  \vspace{2mm}

  {\bfseries Submitted:} 06 Dec. 2019\\
  {\bfseries Published:} 21 Feb. 2020\\

  \vspace{2mm}
  {\bfseries Licence}\\
  Authors of papers retain copyright and release the work under a Creative Commons Attribution 4.0 International License (\href{http://creativecommons.org/licenses/by/4.0/}{\color{linky}{CC-BY}}).
}

\hypertarget{summary}{%
\section{Summary}\label{summary}}

Leadership computing facilities around the world support cutting-edge
scientific research across a broad spectrum of disciplines including
understanding climate change \cite{Kurth_2018}, combating opioid
addiction \cite{Joubert:2018:AOE:3291656.3291732}, or simulating the
decay of a neutron \cite{8665785}. While the increase in computational
power has allowed scientists to better evaluate the underlying model,
the size of these computational projects has grown to a point where a
framework is desired to facilitate managing the workflow. A typical
scientific computing workflow includes:

\begin{enumerate}
\def\labelenumi{\arabic{enumi}.}
\tightlist
\item
  Defining all input parameters for every step of the computation;
\item
  Defining dependencies of computational tasks;
\item
  Storing some of the output data;
\item
  Post-processing these data files;
\item
  Performing data analysis on output.
\end{enumerate}

\href{https://github.com/callat-qcd/espressodb/}{EspressoDB} is a
programmatic object-relational mapping (ORM) data management framework
implemented in Python and based on the
\href{https://www.djangoproject.com}{Django web framework}. EspressoDB
was developed to streamline data management, centralize and promote data
integrity, while providing domain flexibility and ease of use. It is
designed to directly integrate in utilized software to allow dynamical
access to vast amount of relational data at runtime. Compared to
existing ORM frameworks like
\href{https://www.sqlalchemy.org}{SQLAlchemy} or Django itself,
EspressoDB lowers the barrier of access by simplifying the project setup
and provides further features to satisfy uniqueness and consistency over
multiple data dependencies. In contrast to software like
\href{https://github.com/iterative/dvc}{DVC},
\href{https://www.vistrails.org/index.php/Main_Page}{VisTrails}, or
\href{https://taverna.incubator.apache.org}{Taverna}
\cite{10.1093/nar/gkt328}, which describe the workflow of computations,
EspressoDB rather interacts with data itself and thus can be used in a
complementary spirit.

The framework provided by EspressoDB aims to support the ever-increasing
complexity of workflows of scientific computing at leadership computing
facilities (LCFs), with the goal of reducing the amount of human time
required to manage the jobs, thus giving scientists more time to focus
on science.

\hypertarget{features}{%
\section{Features}\label{features}}

Data integrity is important to scientific projects and becomes more
challenging the larger the project. In general, a SQL framework
type-checks data before writing to the database and controls
dependencies and relations between different tables to ensure internal
consistency. EspressoDB allows additional user-defined constraints not
supported by SQL (\emph{e.g.} unique constraints using information
across related tables). Once the user has specified a set of conditions
that entries have to fulfill for each table, EspressoDB runs these
cross-checks for new data before inserting them in the database.

EspressoDB also supports collaborative and open-data oriented projects
by leveraging and extending Django's ORM interface and web hosting
component. The object oriented approach allows the whole team to
determine table architectures without knowing SQL. Once tables have been
implemented by users familiar with the details of the EspressoDB
project, additional users can access data without detailed knowledge of
the project itself. In addition to providing a centralized data
platform, it is possible to spawn customized web pages which can be
hosted locally or on the world wide web\footnote{Depending on the
  configuration, it is possible to provide selected access for multiple
  users on different levels.}. EspressoDB simplifies creating projects
by providing default Django configurations that set up, for example,
connections to the database and webpages to view associated tables. For
example, with the default setting, EspressoDB spawns:

\begin{itemize}
\tightlist
\item
  Documentation views of implemented tables;
\item
  A project-wide notification system;
\item
  Project-specific Python interface guidelines which help writing
  scripts to populate the database;
\item
  Admin pages for interacting with data in a GUI.
\end{itemize}

Further views can be implemented to interact with data and use existing
Python libraries for summarizing and visualizing information. This
allows users to create visual progress updates on the fly and to
integrate the database information to the data-processing workflow,
significantly reducing the human overhead required due to improved
automation.

More details, usage instructions, and examples are documented at
\href{https://espressodb.readthedocs.io}{espressodb.readthedocs.io}.

\hypertarget{use-case}{%
\section{Use case}\label{use-case}}

\href{https://github.com/callat-qcd/lattedb/}{LatteDB}, an application
of EspressoDB, contains table definitions for lattice quantum
chromodynamics (LQCD) calculations and analysis. LatteDB is currently
being used by the \href{https://a51.lbl.gov/~callat/webhome/}{CalLat
Collaboration} in their computations on Summit at the Oak Ridge
Leadership Computing Facility (\href{https://www.olcf.ornl.gov}{OLCF})
through DOE INCITE Allocations \cite{incite:2019,incite:2020}. The
website generated by LatteDB used by CalLat can be found at
\url{https://ithems.lbl.gov/lattedb/}. A precursor to EspressoDB and
LatteDB was used to support a series of LQCD projects
\cite{Nicholson:2018mwc,Chang:2018uxx}.

Summit at OLCF is disruptively fast compared to previous generations of
leadership-class computers. There are two challenges which are both
critical to address for near-exascale computers such as Summit, which
will become more important in the exascale era:

\begin{enumerate}
\def\labelenumi{\arabic{enumi}.}
\tightlist
\item
  \emph{Efficient bundling and management of independent tasks}.
\item
  \emph{Dependent task generation and data processing};
\end{enumerate}

Using lattice QCD as a specific example, the computations can be
organized as a directed multigraph: a single calculation requires
tens-of-thousands to millions of independent MPI tasks to be completed.
These tasks, while running independently, have nested and chained
interdependencies (the output of some tasks are part of the input for
other tasks). Several such complete computations must be performed to
extract final answers. As a specific example, CalLat creates petabytes
of temporary files that are written to the scratch file system, used for
subsequent computations and ultimately processed down to hundreds of
terabytes that are saved for analysis. It is essential to track the
status of these files in real-time to identify corrupt, missing, or
purgeable files.

Understandably, LCFs prohibit the submission of millions of small tasks
to their supercomputers (clogged queues, overtaxed service nodes, etc.).
It is therefore imperative to have a task manager capable of bundling
many tasks into large jobs while distributing the work to various
components of the heterogeneous nodes; To keep the nodes from going
idle, the jobs must be backfilled while running with the next set of
available tasks (item 1). Members of CalLat are addressing the task
bundling through the creation of job management software,
\href{https://github.com/evanberkowitz/metaq}{METAQ}
\cite{Berkowitz:2017vcp}, and \texttt{MPI\_JM}
\cite{8665785,Berkowitz:2017xna}, while LatteDB is designed to address
the dependent task generation. A future feature of LatteDB is
integration with \texttt{MPI\_JM}.

For the second item, keeping track of the tasks, optimizing the order of
tasks and ensuring no work is repeated requires a task manager that
understands all these dependencies and the uniqueness of each task.
Software to track and manage such a computational model at runtime,
which does not require in-depth knowledge of, e.g., managing databases,
does not currently exist, which motivated the creation of EspressoDB and
LatteDB.

\hypertarget{acknowledgements}{%
\section{Acknowledgements}\label{acknowledgements}}

We thank Evan Berkowitz, Arjun Gambhir, Ben Hörz, Kenneth McElvain and
Enrico Rinaldi for useful insights and discussions which helped in
creating EspressoDB and LatteDB. C.K. gratefully acknowledges funding
through the Alexander von Humboldt Foundation through a Feodor Lynen
Research Fellowship. The work of A.W-L. was supported by the Exascale
Computing Project (17-SC-20-SC), a joint project of the U.S. Department
of Energy's Office of Science and National Nuclear Security
Administration, responsible for delivering a capable exascale ecosystem,
including software, applications, and hardware technology, to support
the nation's exascale computing imperative.

This research used resources of the Oak Ridge Leadership Computing
Facility, which is a DOE Office of Science User Facility supported under
Contract DE-AC05-00OR22725, with access and computer time granted
through the DOE INCITE Program.


\bibliography{paper}
\bibliographystyle{ieeetr}

\appendix
\section{Supplemental Information}

\subsection{Django overview\label{django-overview}}

\texttt{EspressoDB} utilizes Python's \texttt{Django} Object-Relational
Mapping (ORM) framework. Tables correspond to Python classes, rows
correspond to instances and columns correspond to attributes of the
instances.  One can filter instances by their attributes
or generate summary tables (\texttt{pandas.DataFrame}) within one line
of code. Furthermore, using an ORM allows one to have the same interface
independent of the backend. It is possible to store data in a file based
\texttt{SQLite} solution, or use more scalable options like
\texttt{MySQL} or \texttt{Postgresql}.

\texttt{Django} is part of many open-source projects and thus comes with
extensive documentation. Additionally, \texttt{Django} is scalable,
comes with reliable tests and vast community support which manifests in
the fact that it is commonly used in large scale projects (BitBucket,
Instagram, Mozilla, NASA and many more). One guiding principle of
\texttt{EspressoDB} is to not ``re-invent the wheel'' but instead
leverage the support coming from \texttt{Django}. As a result, one can
easily incorporate many of \texttt{Django}'s extensions and find
solutions to technical questions online.

\subsection{Lattice QCD use case \label{lattice-qcd-use-case}}

LQCD is an inherently a stochastic method of simulating quantum
chromodynamics (QCD) the fundamental theory of nuclear strong
interactions, which is responsible for confining quarks into protons and
neutrons and ultimately, for binding these nucleons into the atomic
nuclei we observe in nature. The application of LQCD to forefront
research applications in nuclear physics is necessary to build a
quantitative connection between our understanding of nuclei and QCD.
This is important as nuclei serve as laboratories and/or detectors for
many experiments aiming to test the limits of the Standard Model of
particle physics in our quest to understand questions such as: Why is
the universe composed of matter and not anti-matter? Does dark matter
interact with regular matter other than gravitationally? What is the
nature of the neutrino and is it related to the observed excess of
matter over anti-matter? See Ref. \cite{Drischler:2019xuo} for a recent
review of the interface of LQCD with our understanding of nuclear
physics.

The application of LQCD to nuclear physics is an exascale challenge. One
of the main reasons these calculations are so expensive is that when
LQCD is applied to systems of one or more nucleons, an exponentially bad
signal-to-noise problem must be overcome. While the optimal strategy for
overcoming this challenge is not yet known, one thing common to all
methods is the need for an exponentially large amount of statistics. As
such, these LQCD computations require the completion of millions of
independent sub-calculations (or tasks), with chained dependencies, in
order to complete a single calculation. These chained tasks write large
amounts of temporary files to the scratch file system which are used as
input for subsequent files, often with multiple input files required for
the later computations. Several such calculations must be performed in
order to extrapolate the results to the physical point, defined as the
limit of zero discretization (the continuum limit), the infinite volume
limit and the limit of physical quark masses which are \emph{a priori}
unknown and so must be determined through this
extrapolation/interpolation procedure. These requirements lead to a very
complex set of computations that must be carried out with great care and
a significant amount of human time and effort to ensure the computing
cycles are used as efficiently as possible.

Compounding these challenges, the new computer Summit, is
\emph{disruptively fast} compared to previous generations of leadership
class computers. Full machine-to-machine, Summit is approximately 15
times faster than Titan when applied to LQCD applications such as those
carried out by CalLat \cite{8665785}. While this is great for
scientific throughput, it also means the management of the computational
jobs has become unwieldy with the management models typically used for
such computations: Summit churns through the computations so fast, and
produces so much data, it is not possible to keep up with the data
processing and management with our older management tools.

As a concrete example, we consider the nucleon elastic form factor
project being carried out by CalLat \cite{incite:2019, incite:2020}.
For each \emph{ensemble} of gauge configurations used (one choice of
input parameters) the computation requires the following dependent
steps:

\begin{enumerate}
\def\labelenumi{\arabic{enumi}.}
\tightlist
\item
  For each gauge configuration (of O(1000)) in an ensemble (of O(20)),
  make several quark sources (grouped in batches of 8);
\item
  For each source, create a quark propagator;
\item
  For each quark propagator:

  \begin{enumerate}
  \def\labelenumii{\arabic{enumii}.}
  \tightlist
  \item
    Create a proton correlation function to determine the proton mass;
  \item
    Create O(10) proton \emph{sequential} sinks at different separation
    times (times 8 for different spin, flavor and parity combination);
  \end{enumerate}
\item
  For each time-separation, group the 8 \emph{sequential} sinks from the
  8 propagators into a single \emph{coherent sequential} sink
  \cite{Bratt:2010jn};
\item
  For each time-separation, Solve a \emph{sequential} propagator from
  the \emph{coherent sequential} sink;
\item
  For each time-separation, spin, flavor and parity, tie each of the original 8 propagators with
  the sequential propagator to construct a 4-dimensional (4D)
  \emph{formfactor} correlation function;
\item
  Process the data to prepare it for analysis and the extraction of
  physics.
\item
  Repeat for multiple source sets (of 8) if more statistics are
  required.
\end{enumerate}

Each of the steps above leads to the generation of many large
(multi-gigabyte) data files, most of which are not saved.
\texttt{LatteDB} is used to track the status of these data files to know
if they need to be created or if the next step can proceed. The final
data production step, 6, leads to our final data file that need further
processing prior to our final analysis and saving of the files.

Inheriting the functionality of \texttt{EspressoDB}, \texttt{LatteDB}
has the flexibility to faithfully reproduce a one-to-one mapping between
the above computational workflow to database tables. For example, in
step 1, the table of gauge configurations is defined such that every row
in the table specifies a single gauge configuration. This reflects how
on disk, we have thousands of files, each containing a snapshot of the
QCD vacuum, and as such, every file, and every subsequent file as a
result of the gauge configuration (\emph{e.g.} propagators or
correlation functions in steps 2 through 6) can also be tracked
individually. However, at the end of the calculation, an observable is
only well defined with an ensemble of gauge configurations.
\texttt{LatteDB} allows one to define an ensemble table, with a
\texttt{Django\ ManyToMany} data type which may be interpreted as a
single column containing a list of references (foreign keys) to the
table of gauge configurations. In \texttt{SQL}, a list of foreign keys
is not a fundamental data type that is supported, and is only made
possible with \texttt{Django}. However, even with \texttt{Django},
unique constraints can not be implemented on such a column. With
\texttt{LatteDB}, we make it possible to define a unique constraint,
which for this example, prohibits the user from inserting the exact same
list of gauge configurations in the ensemble table more than once. Users
are encouraged to consult the documentation of \texttt{EspressoDB} and
examples in \texttt{LatteDB} for more information.

Additionally, \texttt{LatteDB} is also tremendously useful for managing
the data processing steps (step 7) which proceed as:

\begin{enumerate}
\setcounter{enumi}{6}\item Data processing:
\begin{enumerate}
\item For each configuration, for each time separation, for each of the 8 sources, \emph{time slice} the 4D data \texttt{formfactor} files to only the time slices containing relevant data;

\item For each configuration, for each time separation, for each source set, average the 8 \texttt{formfactor\_tsliced} files to a source averaged file;

\item When all configurations are complete, concatenate these files together;

\item As needed, Fourier Transform these files to project the \texttt{formfactors} onto definite momentum transfers;

\item Analyze the correlation functions to extract the relevant physics.
\end{enumerate}
\end{enumerate}

In Figure 1, we show an example \texttt{LatteDB} Table from step 7b. The
user is able to filter on each column to very quickly assess the
production status (green means the \texttt{tsliced\_source\_averaged} file
exists, red means it is missing) and decide what configurations in what
ensembles need to be managed in production.

\begin{figure}[t!]
\centering
\includegraphics{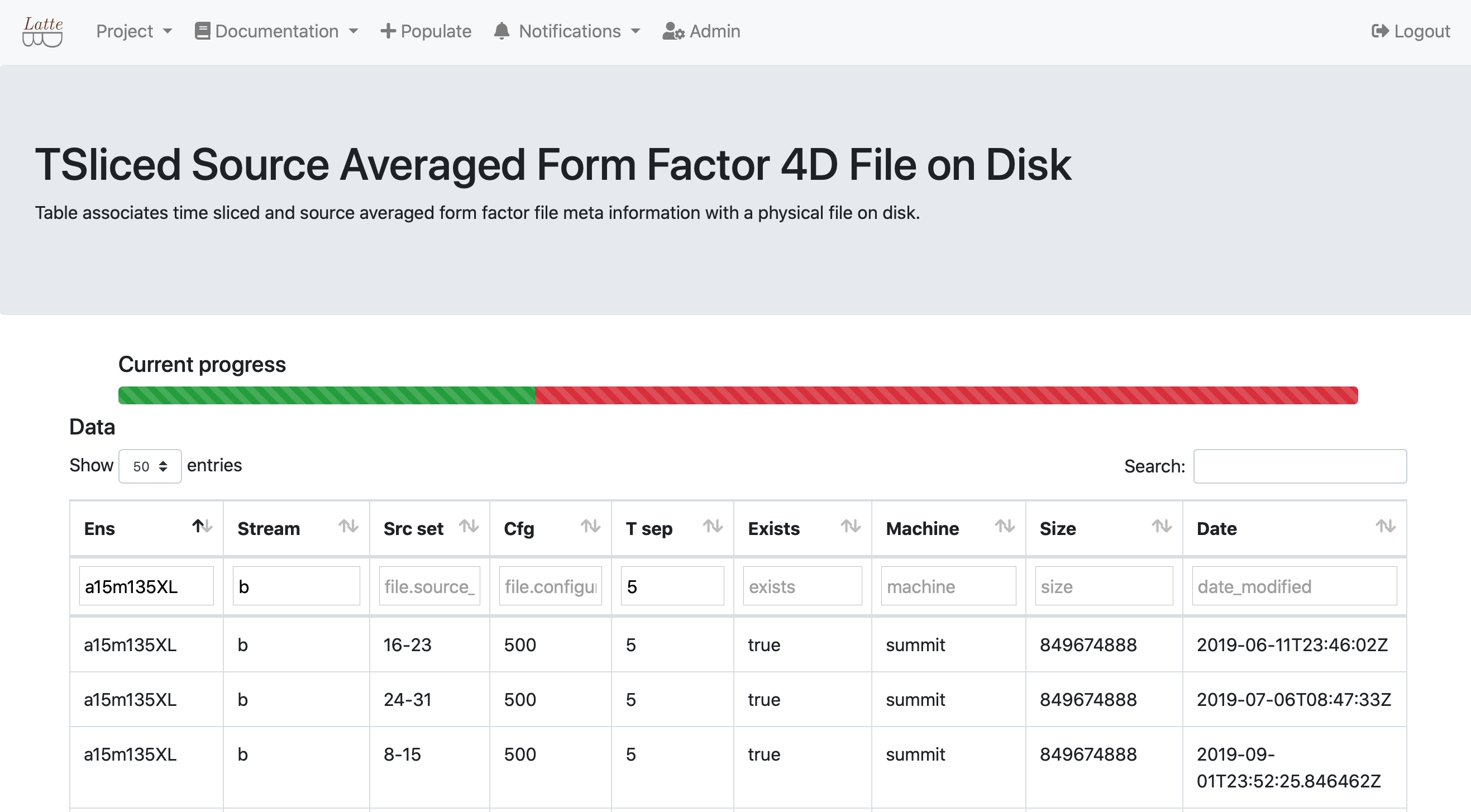}
\caption{Example table view of file status with column specific filters
and dynamic progress bar.}
\end{figure}

More importantly, our production scripts interact with \texttt{LatteDB},
therefore even without the visualization, the scripts will only generate
work that \texttt{LatteDB} records as missing. This interaction with
\texttt{LatteDB} significantly reduces the amount of human time required
to manage the computations. We are actively constructing routines to
also store the final data files to tape, the status of which is stored
in a related \texttt{LatteDB} table. Thus, the user can query the
database instead of the file system for the existence of data files,
significantly reducing the load on the file system as well.
Examples of interacting with \texttt{LatteDB} can be found in
our management repository for these INCITE projects
\url{https://github.com/callat-qcd/nucleon_elastic_FF}.
The scripts for interacting with \texttt{LatteDB} are in the \texttt{scripts} folder and contain \texttt{lattedb} in the name.
Other examples can be found in the \texttt{notebooks} folder in
\texttt{LatteDB}. These scripts will be updated regularly to encompass
more and more utilities from \texttt{EspressoDB} providing a complete
working example.

Other features that can be implemented is the storing of the data files
in \texttt{LatteDB} as well as storing the analysis of the data files.
This allows for communal data analysis within a collaboration with a
centralized location for results, making it easier to combine results
from different members and reduce redundant work. Depending upon the
success and popularity of \texttt{EspressoDB}, it may be worth exploring
whether OLCF (or other LCF) would be willing to allow users to host
databases locally on the machines such that the compute nodes could
interact with the database allowing users to minimize the number of
small input files that are typically written to the file system as well.
In our case, each separate task requires an input file and typically
generates two or three small output files, rapidly polluting the file
system with millions of small files. \texttt{EspressoDB} will minimally
allow users to \emph{clean up} these small files and store the relevant
log and output information in the database.

\end{document}